# Anomalous, Dielectrophoretic Transport of Molecules in Non-Electrolytes

Gaurav Anand,[1] Samira Safaripour,[1] and Craig Snoeyink[1]

[1]Department of Mechanical and Aerospace Engineering, University at Buffalo, Buffalo

Running Title: Anomalous Transport in Non-Electrolytes

Corresponding author: Dr. Craig Snoeyink, 211 Bell Hall, University at Buffalo, Buffalo, NY 14260, USA, Email: craigsno@buffalo.edu

## Non-Standard Abbreviations:

| | | | |
|---|---|---|---|
| AC | Alternating Current | DEP | Dielectrophoretic |
| E-field | Electric Field | FITC | fluorescein isothiocyanate |
| FTO | Fluorine doped Tin Oxide | HER | High Electric Field Region |
| IPA | Isopropyl Alcohol | LER | Low Electric Field Region |
| PMMA | Polymethyl methacrylate | RMS | Root Mean Square |

## Keywords:

# Abstract

The electric field dielectric polarization-based separations mechanism represents a novel method for separating solutions at small length scales. An electric field gradient with a maximum strength of 0.4 MV/m applied across a 10 $\mu$m deep channel is shown to increase the concentration inside the low electric field region by ≈ 40% relative to the high electric field region. This concentration change is two orders of magnitude higher than the estimated change predicted using the classical equilibrium thermodynamics for the same electric field. The deviation between the predicted and the experimental results suggests that the change in volumetric electric field energy with solute concentration is insufficient to explain this phenomenon. The study also explores the effect of varying strength of electric field and frequency of supplied voltage on the dielectric polarization-based separation efficiency. While the increase in the former increases the separation efficiency, the increase in the latter reduces the degree of concentration change due to ineffective charging of the electrodes.

# 1  Introduction

Here, we demonstrate a non-electrophoretic, electric field (E-field) polarization-based separation method that filters solutes with an efficiency that is far higher than predicted by equilibrium thermodynamics. The change in polarization properties of a solution in the presence of an E-field modifies the thermodynamic state of the solution, driving it to a new equilibrium with different concentrations in the high and low electric field regions. Changes in the thermodynamic state due to electromagnetic field polarization effects have long been a subject of study.[1,2] For example, the effect of changes in dielectric properties on a system's equilibrium temperature and pressure in the presence of an E-field has been studied extensively.[3–7] However, to our knowledge, there have been no successful thermodynamic modeling efforts on studying the effect of E-field polarization on the concentration of solutions.

In contrast to polarization-based separations, numerous electrophoretic-based separation methods have been conceptualized and developed utilizing charge transfer to perform separations on solutions. Electrophoresis is the transportation of ions along E-field lines, [8,9] where ions have varying electromobility due to their different sizes and charges. This property has been utilized to separate different species from a solution as they are transported along an E-field in a capillary.[10] Another example is Electrodeionization, where electrophoretic force pulls opposite-charged ions towards electrodes, and resins are used to retain the ions.[11]

The separations method shown here appears to behave similarly to the dielectrophoretic (DEP) transport of particles, which is also a polarization-based mechanism where particles are transported along the gradient of the E-field with direction depending on the dielectric properties of the particles.[12,13] More importantly, dielectrophoresis does not require direct electrical contact with the solution or consumables like filter media.[14] While dielectrophoresis is a well-known particle transportation method, it scales with the cube of the particle radius and has been reported to be hard to utilize for nanoscale molecular transportation.[15] However, we report here, for the first time, that the dielectric polarization-based mechanism works very effectively for sub nanoscale particles too.

To our knowledge, all reported field-based molecular separation methods rely on electrophoretic effects except possibly for recent works on the trapping of large proteins by E-field gradients.[16–21] This DEP trapping of large proteins has been modeled utilizing the DEP transport models of nanoscale particles and those models have largely failed to present a clear physics and an effective model to describe the anomalous separation efficiency for both nanoscale and sub-nanoscale size molecules. The DEP models attempt to treat the proteins as small particles and have underestimated the strength of the transport, sometimes strikingly so. Attempts to reconcile the difference between experiment and models by incorporating increased charge transport along the surface of the proteins were partially successful, but large discrepancies with experiments still exist. [22–25]

In contrast to the "top-down" approach of using particle-based models of DEP transport, this work hypothesizes that a chemical-thermodynamic approach could provide an alternative modeling. To test this hypothesis, we utilized a much smaller solute - fluorescein isothiocyanate (FITC) solvated in methanol - than has been used in the literature to eliminate the contribution of mobile charges on the transport. As we will show, even in this system the equilibrium thermodynamic model developed here drastically underestimates the resulting separations efficiency. The rejection of this hypothesis is a notable finding which suggests that there are physics which are not accounted for in current approaches. This work is not the first to examine field-induced transport in a FITC-methanol system [26] however the experimental system presented here varies from that of An and Minerick in the electrode geometry, electrode insulation, and flow conditions. By utilizing insulated electrodes in a comparatively simple "parallel plate" electrode arrangement we have eliminated the possibility of charge-transfer transport mechanisms while lowering uncertainty regarding the E-field strength and gradients.

# 2 Materials and Methods

## 2.1 Channel fabrication

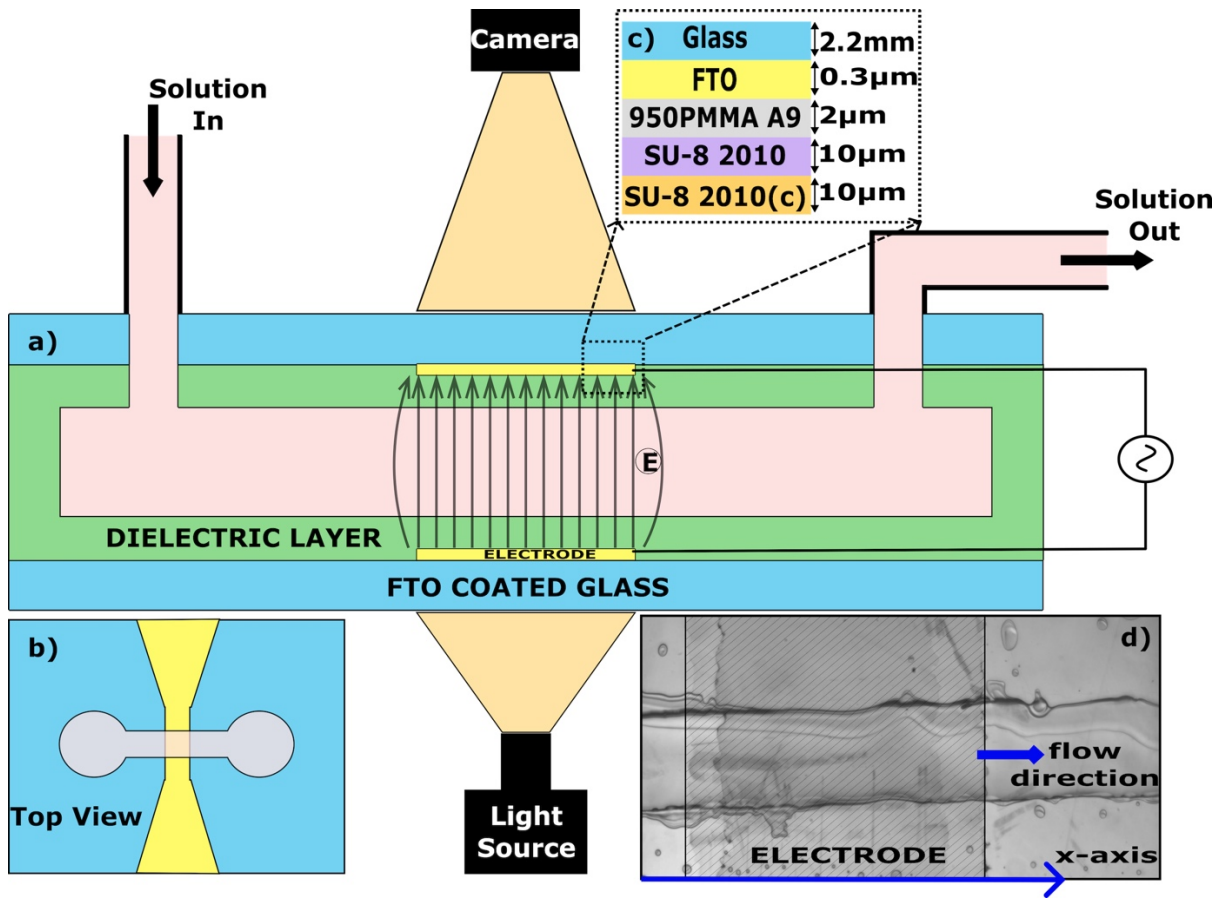

Figure 1: (a) The schematic of the experimental setup. The bottom and top electrodes are connected to an AC power source. (b) Top view showing channel and electrode geometry. (c) Channel wall cross-section showing layers, composition, and thickness (d) An original image of the channel, the hatched area showing the overlapping electrodes.

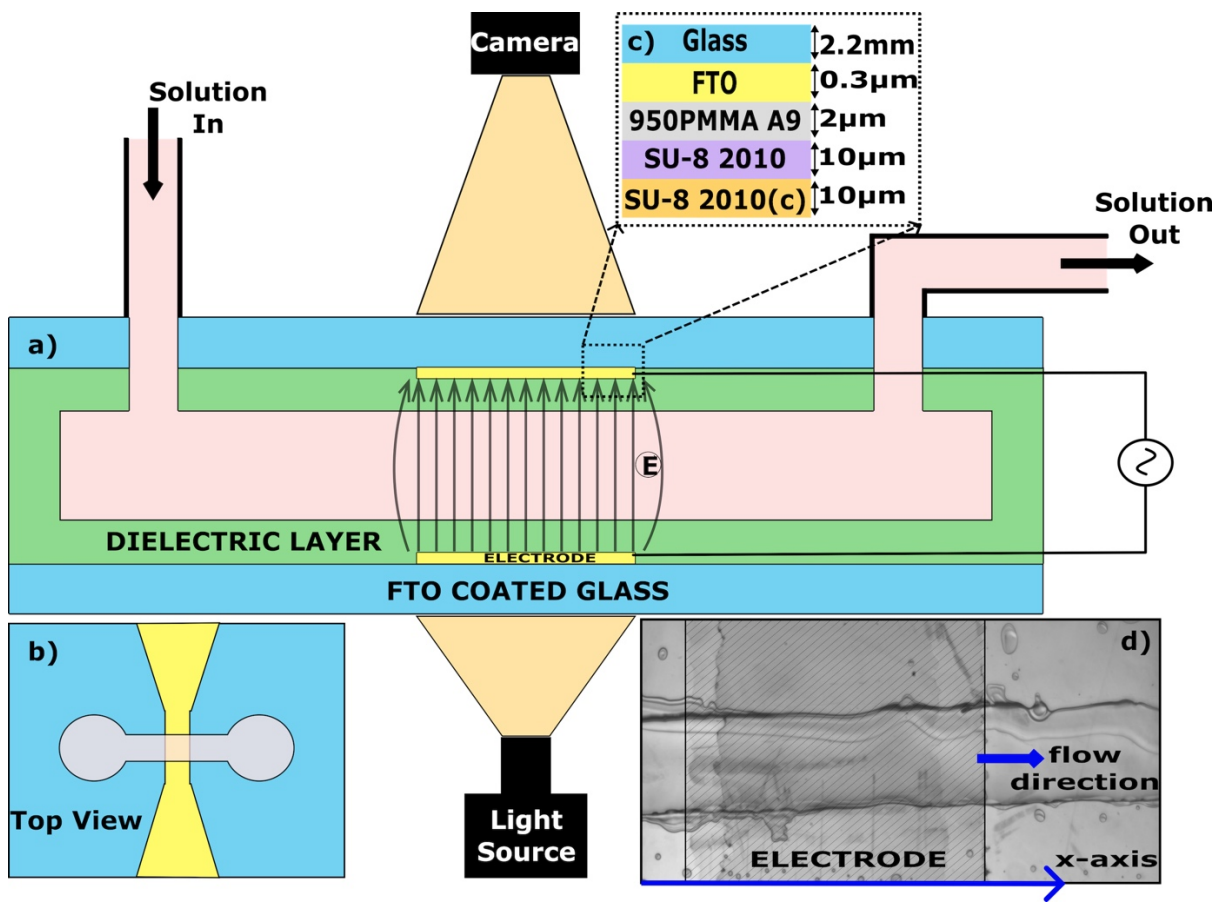

Figure 1 is the schematic of the experimental setup used to study the dielectric polarization-based effect of an externally applied Alternating Current (AC) E-field on a FITC (Fluorescein 5(6)-isothiocyanate)-methanol solution of an initial concentration 0.1 mM, which is confined between two electrodes in a microfluidic channel. The fabrication of the microfluidic channel is based on the principle of photolithography discussed below.[27] The FITC-methanol solution is prepared by mixing Fluorescein 5(6)-isothiocyanate, ≥ 90%, (supplied by Sigma-Aldrich) in Methanol, anhydrous ≥ 99.8%, (supplied by Macron Fine Chemicals). The FITC-methanol solution was pumped across the channel with a syringe pump at a fixed volumetric rate of 180 $\mu$l/hr resulting in an average velocity of 5 mm/s.

The microfluidic channel was constructed with plate electrodes which were insulated using a multi-layer dielectric approach to reduce the number of pinholes, [28] reducing dielectric breakdown events.[29] FTO (Fluorine doped Tin Oxide)-coated glass slides with an electrode thickness of ∼ 320 nm and surface resistivity of ∼ 7 $\Omega/cm^2$ were used as substrate (supplied by MSE Supplies LLC), and electrodes were formed by wet etch followed by ultrasonic cleaning. Two layers of Polymethyl methacrylate (950 PMMA A9) supplied by Microchem Laboratory were spin-coated at 3500 rpm and baked to form the $\sim 1\ \mu m$ thick base layer of electrode insulation. This PMMA layer has high dielectric strength ($\sim$ 3, 5.8 MV/cm), is optically clear [30] and facilitates SU-8 adhesion to the glass and electrodes.[31]

The second layer of insulation was made with SU-8 2010 (supplied by Microchem Laboratory). SU-8 was coated at 3500 rpm and soft baked at $95°C$ for 3 minutes to form a $\sim 10\ \mu m$ thick layer. SU-8's dielectric strength of $\sim 4.4$ MV/cm, [32] and high chemical resistance with very low permeability to a solution makes it a perfect candidate for the top insulation layer. After soft-baking, SU-8 was exposed to 365 nm ultraviolet (UV) rays that activated the photoacid, followed by another baking at $95°C$ to complete the polymerization of monomers to form a solid, hardened layer [33] at the end of baking. [34] We coated another SU-8 layer with the same coating procedure and a mask-negative exposure. The second layer was developed and cleaned using SU-8 developer and Isopropyl Alcohol (IPA) to form the channel on the bottom slide of the chip.

For the top slide, a drilled FTO-coated glass coated with two layers of PMMA and one layer of cured SU-8 utilizing the same procedure as before, was used to seal the channel with the SU-8 layer acting as an adhesive. After alignment, both top and bottom slides were placed in a heated press and held at a temperature of $70°C$ and a pressure of 140 psi for 20 mins, then allowed to cool down to room temperature while keeping the pressure fixed. Finally, the entire chip was exposed to 365 nm UV light to permanently bond the two halves. [35–37] After sealing the channel, an AC power supply was attached to the top and bottom slide via high-voltage electrodes, and the solution was supplied with a micro pump attached to the inlet and outlet of the channel. To estimate the E-field strength inside the solution, the system was modeled as a series of capacitors corresponding to the layers of PMMA, SU-8, and the solution. Using this model, an applied potential drop of 75 V across electrodes would result in a voltage drop of 2 V and an effective E-field strength of 0.2 MV/m across the 10 $\mu m$ channel. We applied an RMS voltage varying from 75 – 150 V, resulting in an E-field strength varying from ~ 0.2 – 0.4 MV/m.

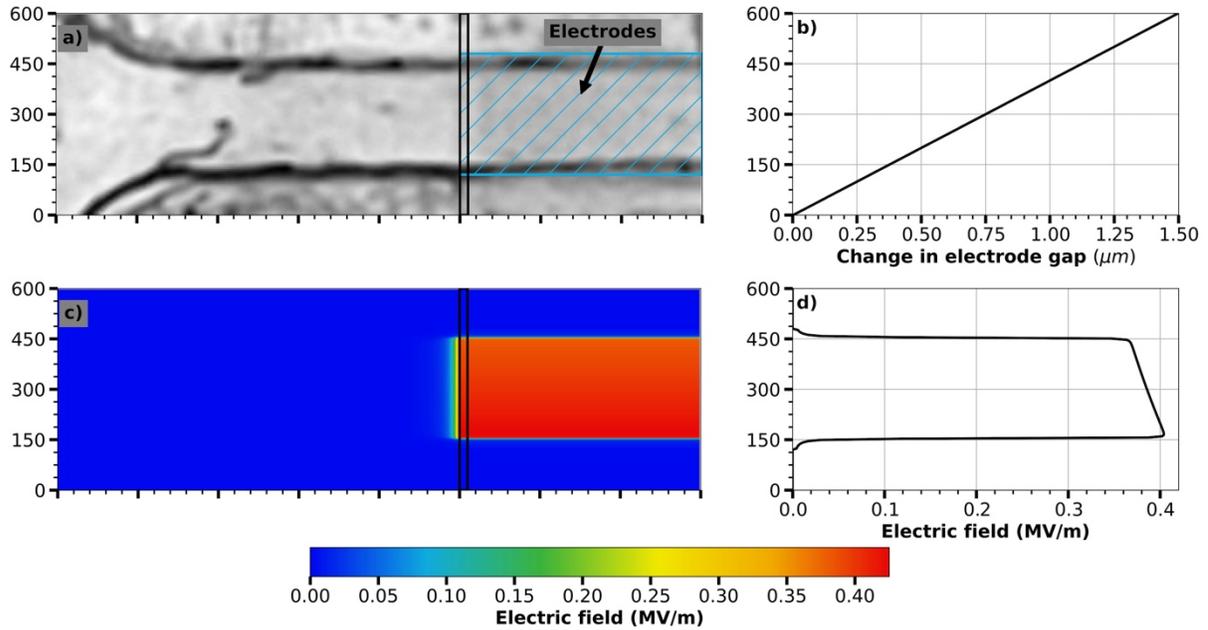

Figure 2: The electrode configuration and the estimated E-field strength. (a) Image of the channel, where the overlapping electrodes are represented by the hatched area. (b) A line plot representing the change in the gap between electrodes perpendicular to the channel measured using a profilometer at a position shown in (a) by a black rectangular box. (c) A contour plot of E-field strength at the center of the channel. (d) A line plot of E-field strength at a position at the center of the channel shown in (c) by a black rectangular box.

The electrodes were configured in a way such that they produce a spatially non-uniform E-field inside the channel. Figure 2 shows the electrode configuration and the estimated E-field strength at the center - midway of the channel's height - of the channel for an applied AC voltage of an amplitude 150 $V_{pp}$ and frequency 10 $kHz$. An approximately 1 $\mu m$ change in electrode gap along the channel's width direction, shown in Figure 2(b), produces a gradient in E-field that varies by an amount of ~ 0.05 MV/m as shown in Figure 2(d). The change in electrode gap along the direction perpendicular to the flow was measured with a stylus-based profilometer (KLA-Tencor Alpha-Step IQ). The electrode gap is due to a difference in the thickness of the SU-8 layer across the channel and not due to a variation in the thickness of the glass. To make sure, the glass substrates' thickness was measured by profilometer before coating any photoresist layers, and no significant difference in the glass thickness across the surface was observed. The estimated strength of the E-field along the centerline of the channel varies from 0 to 0.4 MV/m as it transitions into the E-field region as shown in Figure 2(c).

To estimate the E-field strength inside the solution, the system was modeled as a series of capacitors corresponding to the layers of PMMA, SU-8, and the solution, which have different dielectric constants. The dielectric constant of SU-8 and PMMA varies between 3 and 4, which has been reported before by many researchers. [30,32] For simplicity, we assumed a dielectric constant of 3 for both the dielectric layers, i.e., SU-8 and PMMA. The dielectric constant of the methanol was assumed to be that of the pure methanol due to the low concentration of FITC, which is 32.7. Using this model, an applied potential drop of 75 V across electrodes would result in a voltage drop of 2 V and an effective E field strength of 0.2 MV/m across the 10 $\mu$m channel. We applied an RMS voltage varying from 75 – 150 V, resulting in an electric field strength varying from ~ 0.2 – 0.4 MV/m.

## 2.2 Thermodynamic Model for FITC-Methanol System

To model the separations, we used the change in Helmholtz free energy $(dF')$, developed earlier by the group, [38] under the assumption that no other forces except E-field $(\vec{E})$ is acting on the system as given by Equation 1,

$$dF' = -\left(S + \frac{E^2 V}{2}\frac{\partial \epsilon}{\partial T}\Big|_{c_j,p,\omega,\vec{E}}\right)dT - \left[p - \frac{E^2}{2}\left(\epsilon + \sum_{i=1}^{k} c_i \frac{\partial \epsilon}{\partial c_i}\Big|_{c_j \neq c_i,p,T,\omega,\vec{E}}\right)\right]dV$$

$$+ \vec{E}V d\vec{D} + \sum_{i=1}^{k}\left(\mu_i - \frac{E^2}{2}\frac{\partial \epsilon}{\partial c_i}\Big|_{c_j \neq c_i,p,T,\omega,\vec{E}}\right)dn_i \quad (1)$$

$$- \frac{E^2 V}{2}\left(\frac{\partial \epsilon}{\partial p}\Big|_{c_j,T,\omega,\vec{E}}dp + \frac{\partial \epsilon}{\partial \omega}\Big|_{c_j,p,T,\vec{E}}d\omega + \frac{\partial \epsilon}{\partial \vec{E}}\Big|_{c_j,p,T,\omega}d\vec{E}\right)$$

where, S represents entropy, E is electric field strength, V is volume, $\epsilon$ is permittivity of the solution, T is temperature, $c_i$ is concentration of species i in molarity, p is pressure, ω is frequency, $\vec{D}$ is electrical displacement vector, $\mu_i$ is chemical potential of species i in absence of $\vec{E}$, and $n_i$ is number of moles of species i. Alongside the assumptions used by Anand et al. [38], we assumed the permittivity of the solution is independent of frequency, pressure, and E-field strength because these parameters are small compared to the magnitudes needed to observe detectable differences in dielectric properties.[39] Additionally, we assume that the term $\vec{E}V d\vec{D} = V\int_0^t (\vec{E} \cdot d\vec{D})dt$ is averaged to zero in the presence of AC at very high frequency, resulting in Equation 2:

$$dF' = -\left(S + \frac{E^2 V}{2}\frac{\partial \epsilon}{\partial T}\Big|_{c_j}\right)dT - \left[p - \frac{E^2}{2}\left(\epsilon + \sum_{i=1}^{k} c_i \frac{\partial \epsilon}{\partial c_i}\Big|_{c_j \neq c_i,T}\right)\right]dV$$

$$+ \sum_{i=1}^{k}\left(\mu_i - \frac{E^2}{2}\frac{\partial \epsilon}{\partial c_i}\Big|_{c_j \neq c_i,T}\right)dn_i \quad (2)$$

The movement of ions in the presence of E-field increases the temperature of the solution due to motion resistance, also known as Joule heating. As shown by Anand et al., [40] the E-field required to raise the temperature with forced convection boundary conditions is well above the voltage applied to the non-conductive solution utilized here. Additionally, we can assume fixed temperature and volume because the system achieves a thermal equilibrium much quicker than chemical equilibrium, [38] and we have chosen a closed system for our analysis. With these assumptions and considering any change in the number of moles of species *i* in high E-field region (HER) is equal to the change in number of moles of species *i* in low E-field region (LER) ($dn_i^{HER} = -dn_i^{LER}$) because of conservation of moles, reduces Equation 2 to Equation 3,

$$0 = \sum_{i=1}^{k} \left( \mu_i^{HER} - \mu_i^{LER} - \frac{\epsilon_0 \Delta E^2}{2} \frac{\partial \epsilon'}{\partial c_i} \bigg|_{c_j \neq c_i, T} \right) d n_i^{HER} \quad (3)$$

where $\mu_i^{HER}$ and $\mu_i^{LER}$ are the chemical potential of species *i* in the HER and LER respectively, $\epsilon_0$ is the permittivity of vacuum, $\epsilon'$ is the dielectric constant of the solution, $\Delta E$ is the difference in E-field strength between HER and LER, and $n_i^{HER}$ is the number of moles of species *i* in HER.

To quantitatively explore this model, we utilized a simple binary solution of methanol and a solute, which in our case is FITC. Since published concentration-based permittivity data is lacking for the FITC-methanol solution, we measured the change in permittivity of FITC-methanol solutions with concentration as −20 $mM^{-1}$ in the dilute regime, utilizing the same experimental setup. We analyzed the current-voltage characteristics using an oscilloscope (PicoScope® 4262, supplied by Pico Technology Ltd.) for different concentrations of this solution to evaluate the change in impedance with concentration using the electrical circuit model discussed above and consequently estimated the change in permittivity with concentration. The change in the dielectric constant of FITC-methanol solutions with concentration is a negative and constant value, which is a consistent result based on observations made by many researchers in a dilute regime. [41,42] For reasonably dilute systems such as that used here the change in permittivity with concentration is well behaved and reasonably modeled as a constant. [41] The change in the number of moles of methanol and solute in HER is then given by Equation 4,

$$0 = \left[\ln\left(\frac{a_s^{HER}}{a_s^{LER}}\right) - \frac{v_s \epsilon_0 \Delta E^2}{2RT}\left(\epsilon' + \sum_{i=s,m} c_i \frac{\partial \epsilon'}{\partial c_i} + \frac{1}{v_s}\frac{\partial \epsilon'}{\partial c_s}\right)\right] dn_s^{HER}$$
$$+ \left[\ln\left(\frac{a_m^{HER}}{a_m^{LER}}\right) - \frac{v_m \epsilon_0 \Delta E^2}{2RT}\left(\epsilon' + \sum_{i=s,m} c_i \frac{\partial \epsilon'}{\partial c_i} + \frac{1}{v_m}\frac{\partial \epsilon'}{\partial c_m}\right)\right] dn_m^{HER} \quad (4)$$

where, $a_s^{HER}$, and $a_s^{LER}$ are mean molal activities of solute in HER and LER respectively, $v_s$, and $v_m$ are the molar volume of solute and methanol respectively, $R$ is the universal gas constant, $n_s^{HER}$, and $n_m^{HER}$ are the number of moles of solute and methanol in HER respectively, and $a_m^{HER}$, and $a_m^{LER}$ are mean molal activities of methanol in HER and LER respectively.

Since the volumetric compressibility, bulk modulus, of solution and the molar volume of species are constant because of fixed temperature and pressure, the change in the number of moles of methanol and solute can be related to their molar volumes according to Equation 5.

$$\sum_{i=1}^{k} v_i d n_i = v_s dn_s + v_m dn_m = 0 \quad (5)$$

At the E-field strengths used here, the change is pressure due to electrostriction is estimated to be on the order of $10^{-4}$% and is assumed to be negligible. [43–47] A relationship between activities of species in HER and LER can then be established after substituting Equation 5 in Equation 4, results in Equation 6.

$$\ln\left(\frac{a_s^{HER}}{a_s^{LER}}\right) - \frac{v_s}{v_m}\ln\left(\frac{a_m^{HER}}{a_m^{LER}}\right) = \frac{\epsilon_0 \Delta E^2}{2RT}\left(\frac{\partial \epsilon'}{\partial c_s} - \frac{v_s}{v_m}\frac{\partial \epsilon'}{\partial c_m}\right) \quad (6)$$

Since the solution is dilute, any changes in methanol's concentration and activity are assumed to be negligible. With this assumption, the Equation 6 reduces to Equation 7, which gives the final relationship between change in concentration of solute in HER and LER.

$$\ln(\gamma_s^{HER} c_s^{HER}) = \frac{\epsilon_0 \Delta E^2}{2RT} \frac{\partial \epsilon'}{\partial c_s} + \ln(\gamma_s^{LER} c_s^{LER}) \qquad (7)$$

## 3  Results and Discussion

Figure 3 shows the percentage change in fluorescent emission intensity over time for 0.1 mM FITC-methanol solution at an applied voltage of amplitude 150 $V_{pp}$ and frequency of 10 kHz. The analysis presented in Figure 3 is based on images that are captured at one frame per second over 180 seconds. Figure 3(b) shows the initial intensity change which is negligible at t = 5s i.e., 15 seconds before turning on the E-field. To better understand the effect of E-field on concentration change, four different regions (A, B, C, D) - two outside the E-field region and two inside the E-field region - of width 100 pixels are selected as shown in Figure 3(a). The hatched area in Figure 3 represents the electrodes. As shown, the averaged % intensity change in these four regions at time t = 5 s is negligible, as shown by flat lines in Figure 3(c). The negligible change in intensity at t = 5 s shows that the temporal fluorescent emission intensity change was not affected due to flow or any other factors when the E-field was not turned on.

Figure 3(d), (e) show the intensity change at time t = 25 s, 5 seconds after turning on the E-field, where the % intensity change is again negligible everywhere except for region D. The region D which lies inside the E-field region experiences a decrease in intensity across the channel width at time t = 25 s, which perhaps suggests transport of FITC molecules from region D to further downstream to attain a new concentration equilibrium. The change in intensity in region D also bolsters that any perturbation in the previous equilibrium should start in a region where the E-field is maximum, which in this case is somewhere downstream. A minute later, after the E-field was turned on, a new quasi-equilibrium state has almost fully developed as shown in Figure 3(f), (g), (h), (i) at times t = 80 s and t = 135 s. The development of two different clear streams in Figure 3(f), (h) - one of higher intensity and another of low intensity with a maximum % intensity change of ±20% respectively in both regions - confirms

the existence of a new quasi-equilibrium state. Here the intensity increases in the LER and the concentration decreases in the region with a HER when compared to Figure 2(c), (d).

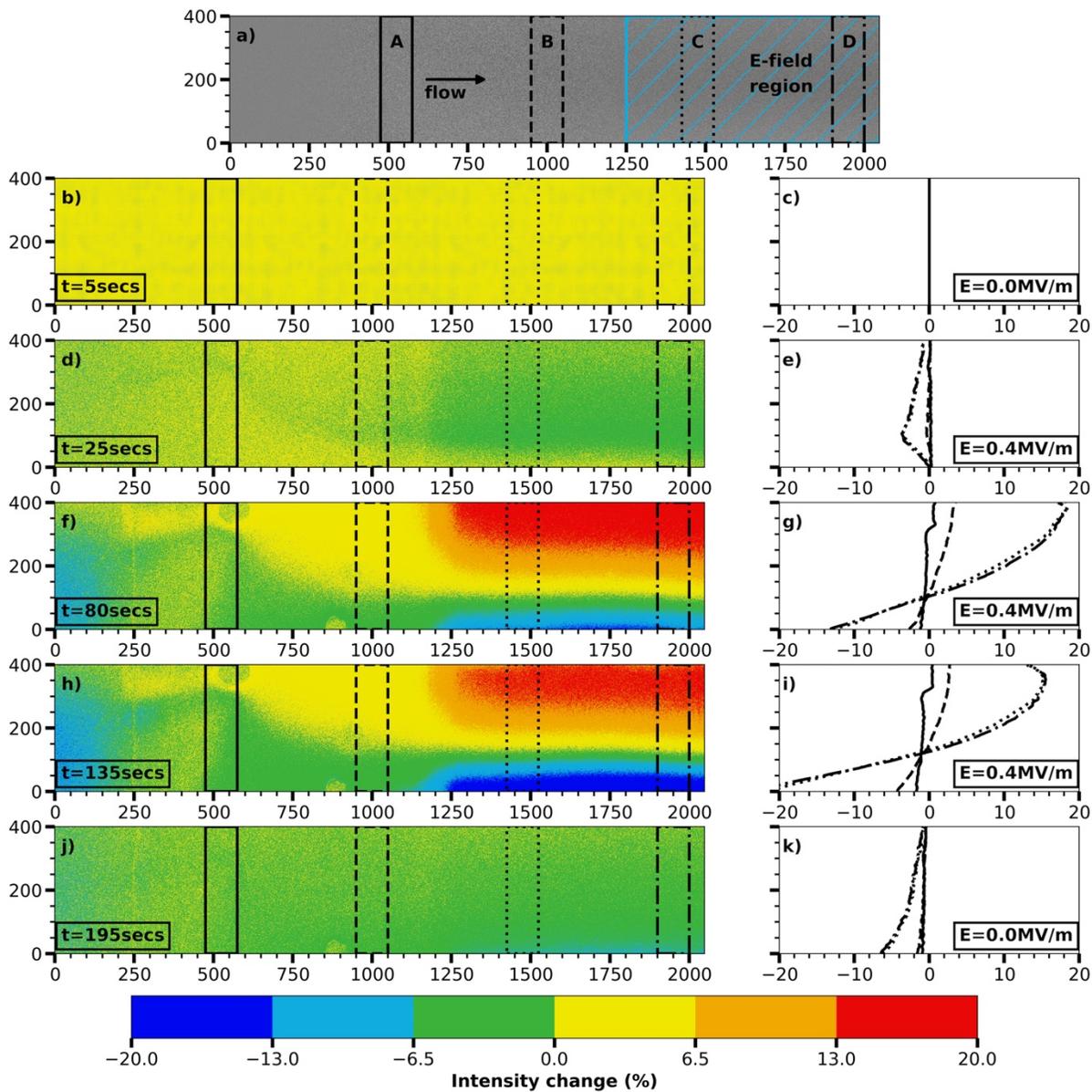

Figure 3: Change in fluorescent emission intensity for 0.1 mM FITC-methanol solution and at an applied voltage of amplitude 150 Vpp and frequency 10 kHz. (a) Image of the channel with an arrow showing the direction of the flow and a hatched area representing the electrodes. (b), (d), (f), (h), (j) shows the contour plots for % change in intensity inside the channel at five

different times varying from t = 5 s to t = 195 s. The E-field was turned on at t = 20 s and turned off at t = 140 s. (c), (e), (g), (i), (k) are the corresponding line plots for % intensity change averaged over 100 pixels at four different regions (A, B, C, D) where region A is shown by the solid line for x = 475 px to x = 575 px, region B is shown by the dashed line for x = 950 px to x = 1050 px, region C is shown by the dotted line for x = 1425 px to x = 1525 px, and region D is shown by the dash-dot line for x = 1900 px to x = 2000 px.

To establish a relationship between the change in fluorescence intensity and dye concentration change, a calibration curve for fluorescence intensity is obtained in section 3.1. Since the maximum % change in intensity observed here is $\pm 20\%$, the change in dye concentration varies linearly with change in intensity, as shown in Figure 6. The separations efficiency of $\approx 20\%$ shown in Figure 3 suggests that the FITC molecules have a strong affinity towards getting transported perpendicular to the flow and along the electric field gradient. While Figure 3 highlights the filtering action of the E-field on the flowing solution in the E-field region, where the E-field depletes FITC molecules from the HER region (y = 0 to y = 200 px) and deposit them into LER (y = 200 to y = 400 px), it also shows that once the quasi-equilibrium state is established, this filtering effect extends well upstream of the electrodes suggesting that even low E-field far from the electrodes are capable of inducing separations.

It is worth noting that the observed separations are not electrophoretic in nature. The FITC solvated in methanol is non-ionic and, as would be expected from an uncharged solute/solvent, no accumulation of fluorescence in the direction of the E-field was observed. If the alternating E-field was driving the solute into the SU-8 walls or causing it to be adsorbed to the channel walls, then we would have observed a slow increase in dye concentration (i.e., a slow increase in fluorescence intensity) within the E-field region as the bulk flow was advecting more dye into this region which then accumulated on the walls. Instead, in all experiments, there was a repeatable transport of solutes from the high to low E-field region and perpendicular to the E-field direction and the resulting separation efficiency reached a steady equilibrium, demonstrating that the ions were not responding electrophoretically to the E-field in any meaningful way.

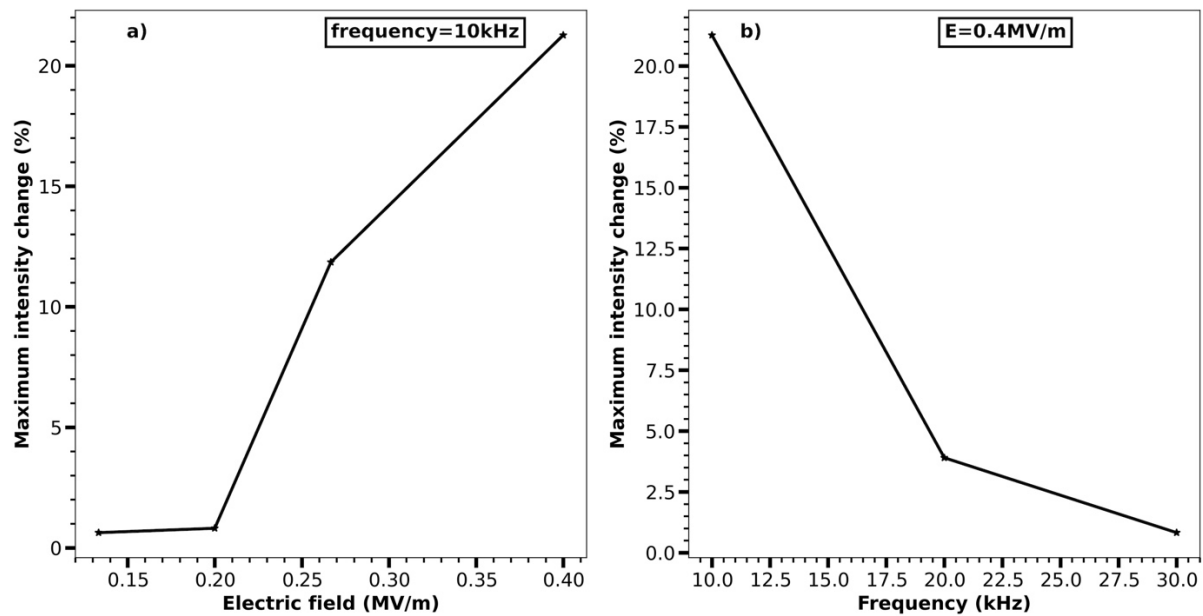

Figure 4: Effect of varying E-field strength and frequency. (a) % intensity change for 0.1 mM FITC-methanol solution as the E-field varies from 0.1 to 0.4 MV/m at a fixed frequency 10 kHz. (b) % intensity change for 0.1 mM FITC-methanol solution as the frequency varies from 10 to 30 kHz at a fixed voltage of 150 Vpp.

The effect of the magnitude of the applied E-field is shown in Figure 4(a), where the maximum fluorescent emission intensity change is plotted against varying E-field strength while the frequency of the supplied voltage is fixed at $10\ kHz$. The effect of the magnitude of the E-field is straight-forward: increasing the applied voltage increases the degree to which dye transports to the low-field region. This effect is as expected from a qualitative examination of Equation 7. The effect of an increase in the frequency of applied voltage is shown in Figure 4(b), where the maximum intensity change % decreases as the frequency increases while the E-field was fixed at $0.4\ MV/m$. Though the applied voltage was fixed in Figure 4(b), an increase in frequency resulted in a smaller effective voltage across the solution due to incomplete charging of the electrodes.

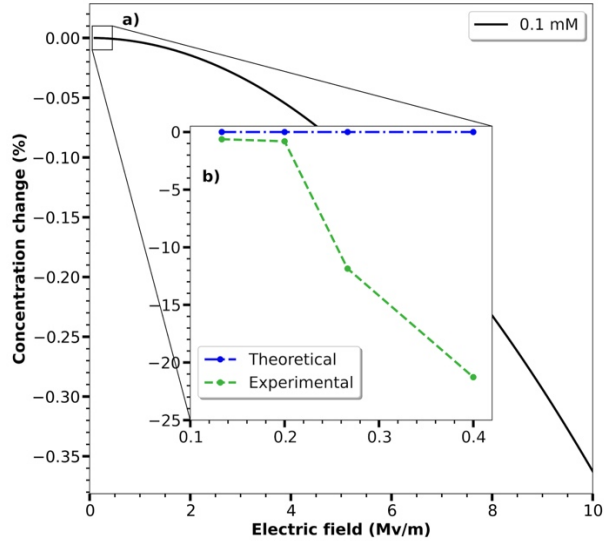

Figure 5: Predicted and measured percent concentration changes as a function of field strength (a) Concentration change predicted by the thermodynamic model for 0.1 mM FITC-methanol solution (b) Difference in concentration change observed with thermodynamic model and the experiment for E-field strength varying from 0.1 – 0.4 MV/m (• represents experimental values).

To compare these results with equilibrium thermodynamics predictions, the equilibrium thermodynamic model Equation 7 for polarization-based separations in a simple binary solution developed above is used. This model qualitatively agrees with our experimental observations. Since the quantity $\partial \epsilon'/\partial c_s$, i.e., change in dielectric constant with concentration is negative for FITC-methanol solutions, equation 7 predicts transport of the solutes to regions with lower E-field-strength exactly as observed in our experiments. Quantitatively, however, the difference between the equilibrium model and experimental observations is stark, as shown in Figure 5(b) where the difference in the estimated and observed concentration change for an E-field range of 0.1 to 0.4 MV/m is substantial. The difference between the observed and estimated change in concentration grows from ∼ 1% at 0.1 MV/m to ∼ 20% at 0.4 MV/m. Equation 7 also predicts that the experimentally observed concentration change should require an E-field strength two orders of magnitude higher, i.e., ∼ 80 MV/m, than utilized in our experiments as shown in Figure 5(a) for a 0.1 *mM* FITC-methanol solution.

This very large difference between equilibrium thermodynamic model predictions and experimental results is notable as the thermodynamic model should provide an upper-bound on the possible separations efficiency. Despite our experimental results being non-equilibrium in nature they exhibit not just a higher separations efficiency, but one that is 20% higher. This is significant because it suggests that crucial physics are being lacking from the equilibrium thermodynamics model; it is unlikely that the assumptions used in the model derivation could produce errors of this magnitude. Unfortunately, the results presented here do not provide direction as to what is missing from the model and as a result future work is warranted.

## 3.1 Calibration curve for fluorescence intensity

Figure 6 shows the calibration curve for fluorescence intensity change with a change in concentration of FITC in methanol. The plot is obtained utilizing the same microfluidic channel. The fluorescence intensity was measured for five different concentrations of FITC-methanol mixture, varying from 0.5 mM to 1.0 mM. The calibration curve from experimental data shown in Figure 6 can be divided into two regimes where for concentration changes up to 20%, the % change in fluorescence intensity varies linearly with the % change in the concentration of the fluorescent dye, and for concentration changes higher than 20%, the % change in fluorescence intensity deviates from the linear behavior, as predicted by Beer-Lambert law. [48] A curve fitted to the experimental data according to Beer-Lambert law is also shown in Figure 6.

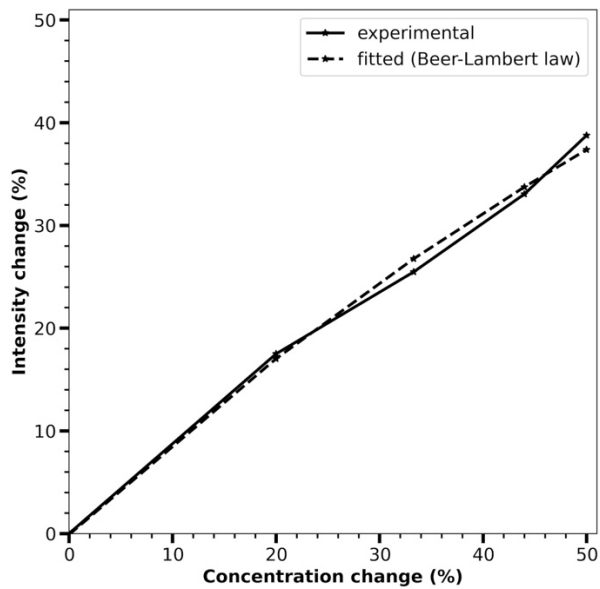

Figure 6: Calibration curve for fluorescence intensity. Black solid line represents the experimental data for % intensity change with % concentration change and black dashed line represents the fitted data using Beer-Lambert law.

## 4  Concluding Remarks

E-fields have long been expected to be capable of altering solution's equilibrium through polarization effects but, to the author's knowledge, have not been experimentally validated. Here an equilibrium thermodynamic model is developed to estimate the separations efficiency under an imposed E-field which is then validated by experimental results. While the model and experiments qualitatively agree, there are substantial quantitative differences where the model underestimates the concentration change by a factor of 20%. The magnitude of the discrepancy between the model and the experiment suggests that the volumetric E-field energy term in Equation 7 is not solely responsible for the observed behavior. While these experiments highlight the deficiency of current theory, they do not suggest what physics could be contributing to the increased separation efficiency.

As a result, further study of this unique phenomenon is warranted, given the surprising strength of the transport. The separation efficiency demonstrated here, on the order of 20%, is significant for microfluidic applications. The small length scales involved make it possible to apply this separations mechanism serially to generate larger concentration changes while its ability to transport both charged and uncharged species makes it uniquely flexible. As a result, it could see an application in lab-on-a-chip medical diagnostics or other sensing applications where space and solution volume are at a premium and where the charge-agnostic nature of the transport is attractive.

# Acknowledgement

The authors thank to the donors of the American Chemical Society Petroleum Research Fund for partial support of this research.

# Conflict of interest statement

All Authors declare that there are no conflicts of interest.

# References

[1] Frank, H. S., Thermodynamics of a Fluid Substance in the Electrostatic Field. *The Journal of Chemical Physics* 1955, 23, 2023–2032.

[2] Zimmels, Y., Thermodynamics in the presence of electromagnetic fields. *Physical Review E* 1995, 52, 1452–1464.

[3] Melcher, J. R., Continuum Electromechanics. MIT Press, Cambridge, Mass 1981.

[4] Cañas-Marin, W. A., Ortiz-Arango, J. D., Guerrero-Aconcha, U. E., Lira-Galeana, C., Thermodynamics of wax precipitation under the influence of magnetic fields. *AIChE Journal* 2006, 52, 2887–2897.

[5] Danielewicz-Ferchmin, I., Banachowicz, E., Ferchmin, A. R., Water phases under high electric field and pressure applied simultaneously. *Journal of Molecular Liquids* 2007, 135, 75–85.

[6] Safaripour, S., Anand, G., Snoeyink, C., Thermodynamic Study of the Electric Field Effect on Liquid–Vapor Mixture at Equilibrium: An Analysis on a Water–Ethanol Mixture. *J. Phys. Chem. B* 2023, DOI: 10.1021/acs.jpcb.3c01578.

[7] Safaripour, S., Anand, G., Snoeyink, C., Thermodynamic Analysis of Capillary and Electric Field Effects on Liquid–Vapor Equilibrium: A Study on the Water–Ethanol Mixture. *J. Phys. Chem. B* 2023, DOI: 10.1021/acs.jpcb.3c05345.

[8] Ramos, A., Morgan, H., Green, N. G., Castellanos, A., Ac electrokinetics: a review of forces in microelectrode structures. *J. Phys. D: Appl. Phys.* 1998, 31, 2338–2353.

[9] Ou, X., Chen, P., Huang, X., Li, S., Liu, B.-F., Microfluidic chip electrophoresis for biochemical analysis. *Journal of Separation Science* 2020, 43, 258–270.

[10] Burggraf, N., Manz, A., Verpoorte, E., Effenhauser, C. S., Michael Widmer, H., de Rooij, N. F., A novel approach to ion separations in solution: synchronized cyclic capillary electrophoresis (SCCE). *Sensors and Actuators B: Chemical* 1994, 20, 103–110.

[11] Pan, S.-Y., Snyder, S. W., Ma, H.-W., Lin, Y. J., Chiang, P.-C., Development of a Resin Wafer Electrodeionization Process for Impaired Water Desalination with High Energy Efficiency and Productivity. *ACS Sustainable Chem. Eng.* 2017, 5, 2942–2948.

[12] Pesch, G. R., Du, F., A review of dielectrophoretic separation and classification of non-biological particles. *ELECTROPHORESIS* 2021, 42, 134–152.

[13] Pethig, R., Review—Where Is Dielectrophoresis (DEP) Going? *J. Electrochem. Soc.* 2017, 164, B3049–B3055.

[14] Castellanos, A., Ramos, A., González, A., Green, N. G., Morgan, H., Electrohydrodynamics and dielectrophoresis in microsystems: scaling laws. *J. Phys. D: Appl. Phys.* 2003, 36, 2584–2597.

[15] Kim, D., Sonker, M., Ros, A., Dielectrophoresis: From Molecular to Micrometer-Scale Analytes. *Anal. Chem.* 2019, 91, 277–295.

[16] Nakano, A., Chao, T.-C., Camacho-Alanis, F., Ros, A., Immunoglobulin G and bovine serum albumin streaming dielectrophoresis in a microfluidic device. *ELECTROPHORESIS* 2011, n/a-n/a.

[17] Staton, S. J. R., Jones, P. V., Ku, G., Gilman, S. D., Kheterpal, I., Hayes, M. A., Manipulation and capture of Aβ amyloid fibrils and monomers by DC insulator gradient dielectrophoresis (DC-iGDEP). *Analyst* 2012, 137, 3227.

[18] Sanghavi, B. J., Varhue, W., Chávez, J. L., Chou, C.-F., Swami, N. S., Electrokinetic Preconcentration and Detection of Neuropeptides at Patterned Graphene-Modified Electrodes in a Nanochannel. *Anal. Chem.* 2014, 86, 4120–4125.

[19] Nakano, A., Camacho-Alanis, F., Ros, A., Insulator-based dielectrophoresis with β-galactosidase in nanostructured devices. *Analyst* 2015, 140, 860–868.

[20] Kikkeri, K., Kerr, B. A., Bertke, A. S., Strobl, J. S., Agah, M., Passivated-electrode insulator-based dielectrophoretic separation of heterogeneous cell mixtures. *Journal of Separation Science* 2020, 43, 1576–1585.

[21] Alazzam, A., Mathew, B., Alhammadi, F., Novel microfluidic device for the continuous separation of cancer cells using dielectrophoresis. *Journal of Separation Science* 2017, 40, 1193–1200.

[22] Nakano, A., Ros, A., Protein dielectrophoresis: Advances, challenges, and applications. *ELECTROPHORESIS* 2013, 34, 1085–1096.

[23] Pethig, R., Limitations of the Clausius-Mossotti function used in dielectrophoresis and electrical impedance studies of biomacromolecules. *ELECTROPHORESIS* 2019, 40, 2575–2583.

[24] Hayes, M. A., Dielectrophoresis of Proteins: Experimental Data and Evolving Theory. *Anal Bioanal Chem* 2020, 412, 3801–3811.

[25] Zavatski, S., Bandarenka, H., Martin, O. J. F., Protein Dielectrophoresis with Gradient Array of Conductive Electrodes Sheds New Light on Empirical Theory. *Anal. Chem.* 2023, 95, 2958–2966.

[26] An, R., Minerick, A. R., Reaction-Free Concentration Gradient Generation in Spatially Nonuniform AC Electric Fields. *Langmuir* 2022, 38, 5977–5986.

[27] Anand, G., Safaripour, S., Tercovich, J., Capozzi, J., Griffin, M., Schin, N., Mirra, N., Snoeyink, C., A simple electrode insulation and channel fabrication technique for high-electric field microfluidics. *J. Micromech. Microeng.* 2023, 33, 125002.

[28] Zhang, H., Yan, Q., Xu, Q., Xiao, C., Liang, X., A sacrificial layer strategy for photolithography on highly hydrophobic surface and its application for electrowetting devices. *Sci Rep* 2017, 7, DOI: 10.1038/s41598-017-04342-z.

[29] Zhou, K., Heikenfeld, J., Dean, K. A., Howard, E. M., Johnson, M. R., A full description of a simple and scalable fabrication process for electrowetting displays. *J. Micromech. Microeng.* 2009, 19, 065029.

[30] Zhang, H. Q., Jin, Y., Qiu, Y., The optical and electrical characteristics of PMMA film prepared by spin coating method. *IOP Conf. Ser.: Mater. Sci. Eng.* 2015, 87, 012032.

[31] Estrada, M., Mejia, I., Cerdeira, A., Iñiguez, B., MIS polymeric structures and OTFTs using PMMA on P3HT layers. *Solid-State Electronics* 2008, 52, 53–59.

[32] Melai, J., Salm, C., Smits, S., Visschers, J., Schmitz, J., The electrical conduction and dielectric strength of SU-8. *J. Micromech. Microeng.* 2009, 19, 065012.

[33] Mitri, E., Birarda, G., Vaccari, L., Kenig, S., Tormen, M., Grenci, G., SU-8 bonding protocol for the fabrication of microfluidic devices dedicated to FTIR microspectroscopy of live cells. *Lab Chip* 2014, 14, 210–218.

[34] Kayakuam, SU-8 2000 DataSheet. 2016; https://kayakuam.com/wp-content/uploads/2020/08/KAM-SU-8-2000-2000.5-2015-Datasheet-8.13.20-final.pdf

[35] Carlier, J., Chuda, K., Arscott, S., Thomy, V., Verbeke, B., Coqueret, X., Camart, J. C., Druon, C., Tabourier, P., High pressure-resistant SU-8 microchannels for monolithic porous structure integration. *J. Micromech. Microeng.* 2006, 16, 2211–2219.

[36] Lima, R. S., Leão, P. A. G. C., Piazzetta, M. H. O., Monteiro, A. M., Shiroma, L. Y., Gobbi, A. L., Carrilho, E., Sacrificial adhesive bonding: a powerful method for fabrication of glass microchips. *Sci Rep* 2015, 5, 13276.

[37] S G Serra, A Schneider, K Malecki, S E Huq, W Brenner, A simple bonding process of SU-8 to glass to seal a microfluidic device. 2007, DOI: 10.13140/2.1.2832.2082.

[38] Anand, G., Safaripour, S., Snoeyink, C., Effects of Frequency and Joule Heating on Height Rise between Parallel Electrodes with AC Electric Fields. *Langmuir* 2022, DOI: 10.1021/acs.langmuir.1c02967.

[39] Banachowicz, E., Danielewicz-Ferchmin, I., Static permittivity of water in electric field higher than $10^8$ V m$^{-1}$ and pressure varying from 0.1 to 600 MPa at room temperature. *Physics and Chemistry of Liquids* 2006, 44, 95–105.

[40] Anand, G., Tabalvandani, S. S., Snoeyink, C., Proceeding of 7th Thermal and Fluids Engineering Conference (TFEC). Begellhouse, Las Vegas, NV, USA 2022, pp. 1043–1053.

[41] Gavish, N., Promislow, K., Dependence of the dielectric constant of electrolyte solutions on ionic concentration: A microfield approach. *Physical Review E* 2016, 94, DOI: 10.1103/PhysRevE.94.012611.

[42] Valiskó, M., Boda, D., The effect of concentration- and temperature-dependent dielectric constant on the activity coefficient of NaCl electrolyte solutions. *The Journal of Chemical Physics* 2014, 140, 234508.

[43] Floriano, W. B., Nascimento, M. A. C., Dielectric constant and density of water as a function of pressure at constant temperature. *Braz. J. Phys.* 2004, 34, 38–41.

[44] Karpov, D. I., Medvedev, D. A., Density dependence of dielectric permittivity of water and estimation of the electric field for the breakdown inception. *J. Phys.: Conf. Ser.* 2016, 754, 102004.

[45] Vaitheeswaran, S., Yin, H., Rasaiah, J. C., Water between Plates in the Presence of an Electric Field in an Open System. *J. Phys. Chem. B* 2005, 109, 6629–6635.

[46] Marcus, Y., Electrostriction of Several Nonaqueous Solvents under Ambient Conditions and Solvation Numbers of Ions in Them. *J Phys Chem B* 2016, 120, 9755–9758.

[47] Tuan, D. F.-T., Fuoss, R. M., ELECTROSTRICTION IN POLAR SOLVENTS. I1. *J. Phys. Chem.* 1963, 67, 1343–1347.

[48] Mayerhöfer, T., Pahlow, S., Popp, J., The Bouguer-Beer-Lambert Law: Shining Light on the Obscure. *ChemPhysChem* 2020, 21, DOI: 10.1002/cphc.202000464.

# Supporting Information

# Animation of Observed Intensity Change

An animation of the observed intensity change for 0.1 mM FITC-methanol solution and at an applied voltage of amplitude 150 $V_{pp}$ and frequency 10 kHz is included.

TOC Graphic

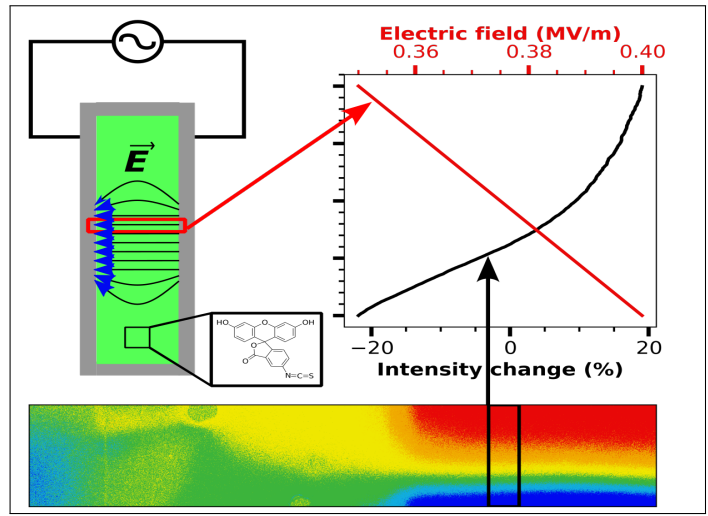